\documentstyle[preprint,tighten,aps,epsfig]{revtex}

\begin{document}
%\draft
%\preprint{HEP/123-qed}
\title{Does the  $D^-/D^+$ production asymmetry decrease at large $x_F$?}
\author{F. Carvalho \thanks{e-mail: babi@if.usp.br}, 
\ F.O. Dur\~aes \thanks{e-mail: fduraes@if.usp.br}, 
\ F.S. Navarra \thanks{e-mail: navarra@if.usp.br}, 
\ M. Nielsen \thanks{e-mail: mnielsen@if.usp.br}} 
\address{Instituto de F\'{\i}sica, Universidade de S\~{a}o Paulo \\ 
C.P. 66318, 05315-970 S\~{a}o Paulo, SP, Brazil}

%\date{\today}
\maketitle
\begin{abstract}
We have applied the meson cloud model (MCM)  to calculate the 
asymmetries in $D$ and $D_s$ meson production in
high energy $\Sigma^-$-nucleus and  $\pi^-$-nucleus 
collisions. We find a good agreement
with  recent data. Our results suggest that the  asymmetries may 
decrease at large $x_F$.
\end{abstract}

\narrowtext

%\section{Introduction}
%\protect\\ The line break was forced via
%$\backslash\backslash$}
%\label{sec:Introdu}

Several experiments have reported \cite{DATA1,DATA2,DATA3,DATA4,WA89_99} 
a significant difference between the  $x_F$ (Feynman momentum) dependence 
of leading and nonleading charmed mesons ($D$). Recent data taken by the 
WA89 collaboration \cite{WA89_99} with a $\Sigma^-$ beam have not only 
confirmed the asymmetry and the leading particle effect but have also 
observed this effect  in $D_s$ and $\Lambda_c$ production. An interesting  
feature of the WA89 data is that they suggest, inspite of very large error 
bars and poor statistics, that the asymmetry decreases at very large  $x_F$. 
More recently, preliminary data from the SELEX collaboration \cite{selex}, 
in contrast to all previous experiments, seem to indicate that the asymmetry  
is smaller and follows a more complicated pattern.

Very soon after the appearence of the first data, it became clear that it 
was not possible to understand them only with usual perturbative QCD  or with 
the string fragmentation model contained in PYTHIA. Alternative models have 
been advanced \cite{nosco,gutierrez}. All these models obtain a 
reasonable description of data, but none of them predicts a decrease in the 
asymmetry. 

The purpose of this letter is to show that in the meson cloud model (MCM) 
\cite{ku} we can reproduce data and accomodate a possible decrease of the 
asymmetry. The MCM has been very successful in the study of hadron structure 
\cite{ku} and of particle production in high energy soft hadron collisions 
\cite{hss,cdnn}. It has been extended to 
the charm sector \cite{nnnt}.

%\section{Asymmetry Production in the Meson Cloud Model}
%\protect\\ The line break was forced via
%$\backslash\backslash$}
%\label{sec:asymmmcm}

In the  MCM we assume that quantum fluctuations in the projectile play an  
important role. Both the  $\Sigma^-$  and $\pi^-$  may be decomposed 
in a series of Fock states. This series has also been discussed in Ref. 
\cite{gutierrez}, where, for example,  fluctuations of the type 
$ |\pi^-\rangle = | \overline{u}d \overline{c} c \rangle$ and 
$ |\Sigma^-\rangle = | d d s \overline{c} c \rangle$ were considered. In the 
MCM we write the Fock decomposition in terms of the equivalent hadronic states 
$| \pi^-\rangle = |D^{0 *} D^-\rangle $ and $|\Sigma^-\rangle = |\Xi^0_c D^-\rangle$. 
This expansion contains  the ``bare'' terms (without cloud 
fluctuations),  light  states and 
states containing the produced charmed meson ($D$ or $D_s$). The latter are, 
of course very much suppressed but they will be responsible for asymmetries. 
The ``bare'' states occur with a higher probability and are responsible for 
the bulk of charm meson production at low and medium momentum ($x_F \leq 0.4$), 
including, for example the perturbative QCD contribution. The cloud states are 
less frequent fluctuations and contribute to $D$ production in the ways described 
below. More precisely we shall assume that: 
\begin{equation}
|\Sigma^-\rangle = Z \,[ \, |\Sigma^-_0\rangle + \, ... \, + \, 
%|M B \rangle \, + \, ...\, + \,      
|\Xi^0_c D^-\rangle + \, |\Sigma^0_c D^-_s\rangle ]
\label{decsig}
\end{equation}
%and 
\begin{equation} 
| \pi^-\rangle = Z' \,[ \, |\pi^0_0\rangle \, + \, ... \, + \,
%| M M' \rangle \, + \, ... \, + \, | B \overline B' \rangle \, + \, ... 
|D^{0 *} D^-\rangle ]
\label{decpi}
\end{equation}
where $Z$ and $Z'$ 
are normalization constants, $|\Sigma^-_0\rangle$ and $|\pi^0_0\rangle$  
are the ``bare'' sigma and pion and the ``dots'' denote possible cloud states 
$|M B \rangle$ in the $\Sigma^-$ and $| M M' \rangle$ or $| B \overline B' \rangle $ 
in the $\pi^-$. The 
relative normalization of these states is fixed once the cloud 
parameters are fixed. We shall first study the reactions induced by 
the cloud component of the $\Sigma^-$. 
This projectile baryon is thus regarded as being a sum 
of virtual meson $(M)$-baryon $(B)$ pairs and a  $\Sigma^-$-proton reaction can thus 
be viewed as a reaction between the ``constituent'' mesons and baryons  of 
the  $\Sigma^-$  with the target proton.

With a $\Sigma^-$ beam the possible reaction 
mechanisms for $D^-$ meson production at large $x_F$ and small $p_T$ (the 
soft regime) are illustrated in Fig. 1.

\begin{figure}
\begin{center}
\epsfig{file=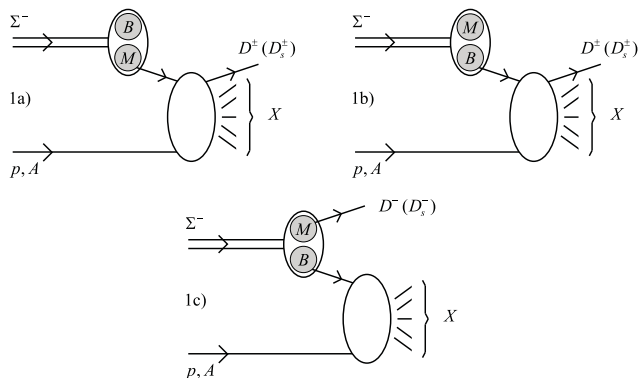,height=5.cm}
\caption{$\Sigma^- p$ collision in which the projectile is in a $|M B\rangle$ state. 
Figs. 1a) and 1b) show the ``indirect'' $D^\pm$ ($D^\pm_s$) production and 1c) the
``direct'' $D^-$ ($D^-_s$) production.}
\label{Fig. 1}
\end{center}
\end{figure}

In Fig. 1a  the baryon  just ``flies 
through'', whereas the corresponding meson interacts inelastically producing 
a $D$ meson in the final state. In Fig. 1b  the meson  just ``flies through'', 
whereas the corresponding baryon interacts inelastically producing a $D$ meson  
in the final state. In Fig. 1c the meson in the cloud is already a $D^-$ (or 
$D^-_s$) which escapes (similar considerations hold for $D^-$  production 
with a $\pi^-$ beam). This last mechanism is responsible for generating 
asymmetries. We shall refer to the  first two processes as  ``indirect 
production'' ($I$) and to the last one as ``direct production'' ($D$). The first 
two are calculated with  convolution formulas whereas the last one is given basically 
by the meson momentum distribution in the cloud initial $ | MB \rangle $ state. 
Direct production has been widely used in the context of the MCM and applied to 
study $n$, $\Delta^{++}$ and $\pi^0$ production \cite{hss}. Indirect meson 
production has been considered previously in  \cite{cdnn}.

Inside the baryon, in the $ | MB \rangle $ state,  the meson and baryon 
have  fractional momentum $y_M$ and $y_B$ with distributions called  
$f_{M/MB}(y_M)$ and $f_{B/MB}(y_B)$ respectively (we shall use for them 
the short notation $f_M$ and $f_B$). Of course, by momentum conservation,  
$y_M + y_B = 1$ and these 
distributions are related by $ f_{M}(y) =  f_{B}(1-y) $ \cite{ku,cdnn}.  
The ``splitting 
function'' $f_{M} (y)$ represents the probability density to find a meson 
with momentum fraction $y$ of the total cloud state $|MB\rangle$. With $f_M$ and 
$f_B$ we can compute the  differential  cross section for production of $D$ mesons, 
which, in the reaction $\Sigma^- p \rightarrow  D X$,  is given by:
\begin{equation}
\frac{ d \sigma^{\Sigma^- p \rightarrow D X}}{d x_F}   =\, \Phi_0 \,\, + \,\,  
\Phi_I\,\, +\,\, \Phi_D
\label{sechoque}
\end{equation}
where $\Phi_0$ and $\Phi_I$  refer respectively to ``bare'' and indirect 
contributions to $D$ meson production and $x_F$ is  the fractional longitudinal 
momentum of the outgoing meson.  $\Phi_D$ represents the direct process depicted 
in Fig. 1c and is given by \cite{hss,cdnn}:
\begin{equation}
\Phi_D =  \frac{\pi}{x_F} \, f_{D} (x_F) \, \sigma^{\Xi}
%\,\, \simeq \,\, \sum_{MB} \frac{ H \,\, f_{M/MB} (x_F )}{x_F} 
\label{direct}
\end{equation}
where $ f_{D} \equiv f_{D^- / \Xi_c^0 D^-}$ and $\sigma^{\Xi}$ is the total 
$p \, \Xi_c^0$ cross section. An analogous expression can be written for 
the reaction $\pi^- p \rightarrow D X$. 

Using (\ref{sechoque}), we can compute the cross sections and also the 
leading ($D^-$)/nonleading($D^+$) asymmetry:
\begin{eqnarray}
A(x_F) &=& \frac{ \frac{d \sigma^{D^-}(x_F)}{d x_F} \,\,-\,\, 
\frac{ d \sigma^{D^+}(x_F)}{d x_F}} {\frac{d \sigma^{D^-}(x_F)}{d x_F} 
\,\, + \,\, \frac{d \sigma^{D^+}(x_F)}{d x_F}}\nonumber \\
&=& \frac{\Phi_D \,\,+\,\,\Phi_I^{D^-}\,\,+\,\,\Phi_0^{D^-}\,\,-\,\,
\Phi_I^{D^+}\,\,-\,\,\Phi_0^{D^+}} {\Phi_D \,\,+\,\,\Phi_I^{D^-}\,\,+
\,\,\Phi_0^{D^-}\,\,+\,\,\Phi_I^{D^+} 
\,\,+\,\,\Phi_0^{D^+}}\nonumber \\
&\simeq& \frac{\Phi_D} {\Phi_D \,\,+\,\, 2 \, \Phi_I^{D}
\,\,+\,\, 2 \, \Phi_0^{D}} \equiv 
\frac{\Phi_D} {\Phi_T} 
\label{assym}
\end{eqnarray}
where the last line follows from assuming $\Phi_I^{D^-}=\Phi_I^{D^+}=\Phi_I^{D}$.   
This last assumption is made just 
for the sake of simplicity. In reality (and also in the calculations performed 
in \cite{nosnafita}) these contributions are not equal and their difference is an 
additional source of asymmetry, which in some cases is not negligible.
Since the ``bare'' states do not give origin to 
$D^- / D^+$ asymmetries (they represent mostly  perturbative QCD contributions
which rarely leave quark pairs in the appropriate kinematic region), 
we have made use of  $\Phi_0^{D+} = \Phi_0^{D-} = \Phi_0^{D}$.  
The denominator of the above expression 
can be replaced by a parametrization of the  experimental 
data:
\begin{equation}
%\Phi_D \,\,+\,\, 2 \, \Phi_I^{D} 
%\,\,+\,\, 2 \,\Phi_0^{D} \,\, = \,\,
\Phi_T = \sigma_0 \, [ \, (1-x_F)^{n^-} \, + \,  (1-x_F)^{n^+} \, ]
\label{sigdata}
\end{equation}
where $n^-$ and $n^+$ are powers used by the different collaborations to fit 
their data and $\sigma_0 \simeq 4 \, - \,7 \,\mu b$ as suggested by the  
data analysis performed in \protect\cite{DATA1,DATA2,DATA3,DATA4,WA89_99,selex}.

Inserting  (\ref{direct}) and (\ref{sigdata}) into (\ref{assym})
the asymmetry becomes:
\begin{equation}
A(x_F) =\frac{\pi \, \sigma^{\Xi}}{ \sigma_0} \,\, 
\frac{ f_{D}(x_F) } {x_F \, [ \, (1-x_F)^{n^-} \, + \,  (1-x_F)^{n^+} \, ]}
\label{finass}
\end{equation}

The behavior of (\ref{finass}) is controlled by $f_D (x_F)$. In the recent works 
with the MCM  one finds two forms for the splitting 
functions. One comes from the evaluation of the relevant Feynman diagrams 
(Sullivan process) \cite{ku,cdnn} and the other comes from a light cone ansatz 
for the cloud state wave function \cite{ma}. We shall compute the asymmetries with 
both of them. The  light cone splitting function is given by \cite{ma}: 
\begin{eqnarray}
f_{M} (y) &=& \sum_{\lambda \lambda'} \,
\int_0^{\infty} \frac{d {\bf k}^2_{\perp}} {16 \pi^2} \, 
| \psi_{\lambda \lambda'}(y,{\bf k}^2_{\perp})|^2 \nonumber \\
&=& \frac{1} {4 \pi^2} 
H^2 \alpha^2 y (1-y) \exp(- \frac{ {\cal M}^2_M}{4 \alpha^2} )
%\end{equation}
%with
%\begin{equation}
%\sum_{\lambda \lambda'} \,  | \psi_{\lambda \lambda'}(y,{\bf k}^2_{\perp})|^2  = 
%A^2  \exp(- \frac{ {\cal M}^2_M}{4 \alpha^2} )
%\label{psi}
\label{fma}
\end{eqnarray}
where $\psi_{\lambda \lambda'}(y,{\bf k}^2_{\perp})$ is the light cone wave function 
of the Fock state containing a meson $M$ (baryon $B$), with longitudinal momentum 
fraction  $y$ (or $1-y$), 
transverse momentum ${\bf k}_{\perp}$ ($- {\bf k}_{\perp}$) and helicity $\lambda$ 
($\lambda'$); $ {\cal M}^2_{M} = \frac{{\bf k}^2_{\perp} + m^2_{M}}{y} \, + \,  
\frac{{\bf k}^2_{\perp} + m^2_{B}}{1-y} $ is the invariant mass of the 
meson ($M$) - baryon ($B$) system for large longitudinal momenta,  
$m_M$ and $m_B$ are their masses,   
$H$ is a normalization constant and $\alpha$ is the width of the heavy meson-baryon 
state.

Substituting (\ref{fma}) into (\ref{finass}) and remembering that in our case 
$y = x_F$, we can  write our final expression  
for the asymmetry:
\begin{eqnarray}
A_{1}(x_F)&=& N_{1} \, \frac{(1 - x_F)}{[ ( 1 - x_F )^{n^-} \, + \, ( 1 - x_F )^{n^+} ]}
\nonumber \\
&\times& \exp \left[ \frac{-1}{4 \alpha^2} \left(\frac{m^2_D}{x_F} + 
\frac{m^2_{\Xi}}{1 - x_F}
\right) \right]
%- \frac{\frac{m^2_D}{x_F} + \frac{m^2_{\Xi}}{1 - x_F}}
%{4 \alpha^2} \right]
\label{finassy}
\end{eqnarray}
where  
%\begin{equation}
$N_{1} = H^2  \, \alpha^2 \, \sigma^{\Xi}/ 4 \, \pi  \, \sigma_0$. 
%\label{defn}
%\end{equation}

A striking feature of (\ref{finassy}) is that in the limit $ x_F \rightarrow 1 $ 
{\it it goes to zero} regardless of the choice of two parameters $N_1$ and $\alpha$.  This 
happens because when the leading meson (responsible for the asymmetry) has momentum 
one its
accompanying  cloud baryon must have momentum zero, being strongly virtual, thereby 
increasing the invariant mass of the cloud state and forcing this cloud configuration 
to have zero probability. For the pion beam, we obtain an analogous expression with the 
replacements $m_{\Xi} \, \rightarrow \, m_{D^{0*}}$, 
$\sigma^{\Xi} \, \rightarrow \, \sigma^{D^{0*}}$. In this case the parameters $H$ and 
$\alpha$ 
may assume different values. Apart from numerical changes, the qualitative behavior of 
$A_{1}$ remains the same, also for large $x_F$.

We now  write the splitting function in the Sullivan process approach. 
The fractional momentum distribution of a pseudoscalar meson $M$  in the state 
$|M B'>$ is given by \cite{ku,cdnn}:
\begin{eqnarray}
f_{M} (y) &=& \frac{g^2_{MBB'}}{16 \pi^2} \, y 
\int_{-\infty}^{t_{max}}
dt \, \frac{[-t+(m_{B'}-m_{B})^2]}{[t-m_{M}^2]^2} \nonumber \\
&\times& F_{MBB'}^2 (t)
\label{ftho}
\end{eqnarray}
where $t$ and $m_{M}$ are the four momentum square and the mass of
the meson in the cloud state and $t_{max} = m^2_B \, y- m^2_{B'} \,  y/(1-y)$ 
is the maximum $t$, with $m_B$ and $m_{B'}$ respectively the $B$ and $B'$ masses. 
Following a phenomenological approach, we use for the baryon-meson-baryon form factor 
$F_{MBB'}$, the exponential form:
\begin{equation}
F_{M B B'} (t) =  \exp  \left( \frac{ t - m_{M}^2}{\Lambda^2_{M B B'}} \right)
\label{eq:form}
\end{equation}
where $\Lambda_{M B B'}$ is the  form factor cut-off parameter. 
%In the above equations $t$ and $m_{M}$ are the four momentum square and the mass of
%the meson in the cloud state and $t_{max}$ is the maximum $t$ given by:

%\begin{equation}
%t_{max} = M^2_B \, y- \frac{M^2_{B'} \,  y}{1-y}  \,\,\,\, , 
%\label{tmax1}
%\end{equation}
%where $M_B$ and $M_{B'}$ are respectively the $B$ and $B'$ masses. 
Considering the
particular case where $B = \Sigma^-$,  $B' = \Xi^0_c$ and $M = D^-$, we   
insert (\ref{ftho}) into
(\ref{finass}) to obtain the final expression for the asymmetry in our second approach:
\begin{eqnarray}
A_{2}(x_F) &=& \frac{N_{2}}{[ ( 1 - x_F )^{n^-} \, + \, ( 1 - x_F )^{n^+} }
\nonumber \\ 
&\times& \int_{-\infty}^{t_{max}}
dt \, \frac{[-t+(m_{\Xi} - m_{\Sigma})^2]}{[t-m_{D}^2]^2}\,
F_{D \Sigma \Xi}^2 (t)
\label{finassytho}
\end{eqnarray}
where 
%\begin{equation}
$ N_{2} = g^2_{D \Sigma \Xi}  \, \sigma^{\Xi} /16 \, \pi  \, \sigma_0$.
%\label{defnp}
%\end{equation}

For the pion beam,  we need also the Sullivan splitting function of the state 
$|\pi^-> \, \rightarrow \, |D^{0 *} D^-\rangle$. In this state,
the $D$ meson  momentum distribution (which was computed in Ref. \cite{nosnafita}) 
turns out to be identical to (\ref{ftho}) except for the
bracket in the numerator which takes the form \cite{nosnafita},  
$[-t + \left( (m_{\pi}^2 - m_{D^{0*}}^2 -t)/2 m_{D^{0*}} \right)^2 ]$, and for 
trivial changes
in the definitions, i.e., $g^2_{MBB'} \, \rightarrow g^2_{\pi D D^{0*}}$, 
$F_{MBB'} \, \rightarrow F_{\pi D D^{0*}}$ and 
$\Lambda_{M B B'} \, \rightarrow \, \Lambda_{\pi D D^{0*}}$.  Realizing that $y = x_F$ 
in the above equations, we can see that in the limit $x_F \rightarrow 1$, $\,$
$t_{max} \rightarrow - \infty$ and the integral in (\ref{finassytho}) goes to zero. In 
fact, it vanishes faster than the denominator and therefore $A_2 \rightarrow 0$. This  
behavior does not depend on the cut-off parameter but it depends on the choice of the 
form factor. For a monopole form factor we may obtain asymmetries which grow even at 
very large $x_F$. Since 
$t$ controls the off-shellness of the emitted meson, which, in turn, is related to the 
virtuality of the $|M B'>$ state (or $|M M'>$ state in the case of the pion beam), the 
vanishing of $A_{2}$ happens for the same physical reason of the vanishing of $A_{1}$.

The $D^-_s / D^+_s$ production asymmetry (with the $\Sigma^-$ beam) can be calculated 
following the steps mentioned above and 
replacing the $|\Xi^0_c D^-\rangle$ state by  $|\Sigma^0_c D^-_s\rangle$. Of course, 
this implies different values for $\alpha$ and $N_{1}$ (in the light cone approach) 
and for $\Lambda$ and $N_{2}$ (in the Sullivan approach), but the qualitative 
discussion (and conclusions) presented above for $A_{1}$ and $A_{2}$ remain valid.

%\section{Results and Discussion}
%\protect\\ The line break was forced via
%$\backslash\backslash$}
%\label{sec:results}

Before presenting our numerical results, we emphasize  that 
i) our calculation is based on quite general and well established ideas, namely 
that hadron projectiles fluctuate into hadron-hadron 
(cloud) states and that these states interact with the target; ii) our results  
only depend on two  parameters: $A_1$ depends on $\alpha$  and  $N_{1}$ and $A_2$ 
depends on $\Lambda$ and $N_{2}$.  Whereas $\alpha$ and $\Lambda$ affect the width 
and position of the maximum of the momentum distribution of the leading  meson in 
the cloud (and consequently of the asymmetry), $N_{1}$ and $N_{2}$ are multiplicative 
constants which determine the strength of the asymmetry.

We show in Figs. 2a and 2b, respectively, our results for 
the asymmetries $D^-/D^+$ and $D^-_s / D^+_s$ (for a $\Sigma^-$ beam). 
The results for a $\pi^-$ beam are shown in Fig. 3. In all figures 
solid (dashed) lines represent $A_{2}$ ($A_{1})$. In both approaches we have two 
parameters  which may be different from reaction to reaction. They are given in 
Tables 1 and 2 below:

%\begin{table}[htb]
\begin{center}
\begin{tabular}{|c|c|c|c|c|c|}\hline
Eq. (\ref{finassy}) & $\alpha \, (GeV)$ & $n^+$ & $n^-$ & $N_{1}$ & 
$\sigma^D/\sigma^T \,(\%)$  \\
\hline
\hline
$\Sigma^- A \rightarrow D^- X$ & $0.77$ & $5.0$ & $4.5$ & $400.00$ & $13.0$ \\
\hline
$\Sigma^- A \rightarrow D^-_s X$ & $0.47$ & $7.0$ & $4.0$ & $0.72 \times 10^9 $ & 
$9.0$ \\
\hline
$\pi^- A \rightarrow D^- X$ & $1.20$ & $5.0$ & $3.5$ & $1.10$ & $8.0$ \\
\hline
\end{tabular}
\end{center}
%\end{table}
\begin{center}
Table 1. {\small Parameters used in the asymmetry $A_{1}$}
\end{center}

%\begin{table}[htb]
\begin{center}
\begin{tabular}{|c|c|c|c|c|c|}\hline
Eq. (\ref{finassytho}) & $\Lambda  \, (GeV)$  & $n^+$ & $n^-$ & $N_{2}$ &
$\sigma^D/\sigma^T \,(\%) $  \\
\hline
\hline
$\Sigma^- A \rightarrow D^- X$ & $2.64$ & $5.0$ & $4.5$ & $2.40$ & $52.0$ \\
\hline
$\Sigma^- A \rightarrow D^-_s X$ & $2.52$ & $7.0$ & $4.0$ & $5.20$ & $67.0$ \\
\hline
$\pi^- A \rightarrow D^- X$ & $2.88$ & $5.0$ & $3.5$ & $0.82$ & $20.0$ \\
\hline
\end{tabular}
\end{center}
%\end{table}
\begin{center}
Table 2. {\small Parameters used in the asymmetry $A_{2}$}
\end{center}

\begin{figure}
\begin{center}
\epsfig{file=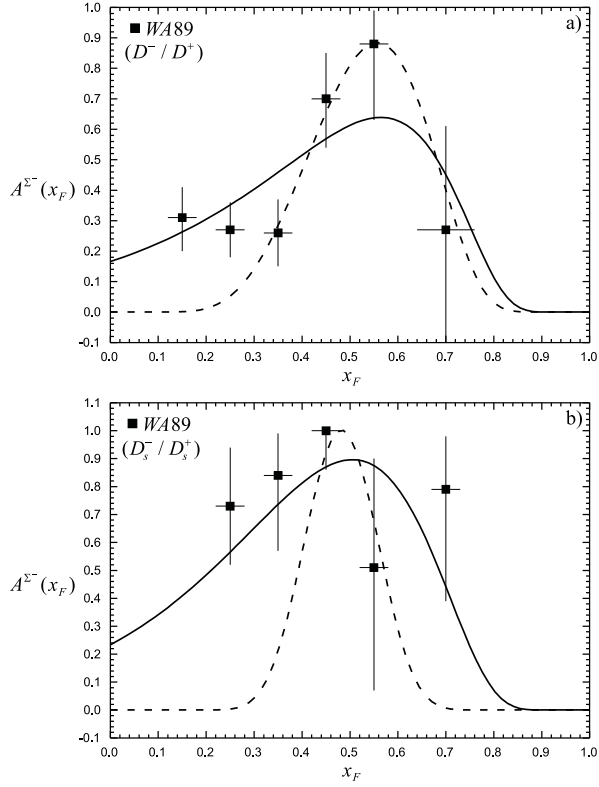,height=10.5cm}
\caption{(2a) Comparison of the MCM asymmetry, Eq.~(\protect\ref{finass}), 
with experimental data \protect\cite{WA89_99} for $D^-/D^+$; (2b) the same as (2a) 
for $D^-_s/D^+_s$.}
\label{Fig. 2}
\end{center}
\end{figure}

\begin{figure}
\begin{center}
\epsfig{file=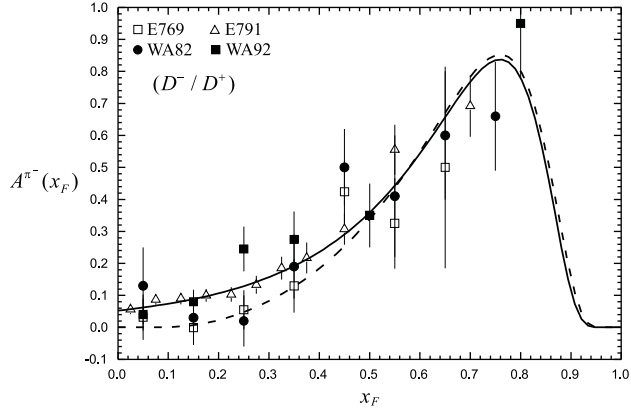,height=5.5cm}
\end{center}
\caption{Comparison of the MCM asymmetry, Eq.~(\protect\ref{finass}), 
with experimental data \protect\cite{DATA1,DATA2,DATA3,DATA4} for $D^-/D^+$.}
\label{Fig. 3}
\end{figure}

As it can be seen in Figs. 2 and 3 the agreement between the MCM and data is very good.
In our picture it is simple to understand why the  $D^-/D^+$ asymmetry peaks at 
$x_F \simeq 0.55$ for the $\Sigma^-$ beam (Fig. 2a) whereas it peaks at a much 
larger value $x_F \simeq 0.8$ for the $\pi^-$ beam (Fig. 3). The $D^-$ meson in the 
$\Sigma^-$ beam originates from the $|\Xi^0_c D^-\rangle$ state and in the $\pi^-$ beam 
it comes from the $| D^{0*} D^-\rangle$ state. Since in the meson-meson state the 
masses are closer than they are in the baryon-meson state, the   $D^-$ is ``faster'' 
inside the pion than inside the $\Sigma^-$. In (\ref{finass}), $f_D(x_F)$ will peak sooner 
and die faster for the $\Sigma^-$. We emphasize that {\it what shifts the peaks of the 
asymmetries are the masses involved} rather than the $\alpha$ (or $\Lambda$) 
parameter. This makes the overall behavior of the asymmetries weakly dependent on 
parameter choices. 

The last column in both tables shows the ratio between the 
direct and total $D^-$ cross sections (where $\sigma^T = \sigma_0 \int dx_F \, 
(1-x_F)^{n^{-}}$). We find this quantity quite model dependent. 
This is also a consistency check. In order to 
treat the cloud as a perturbation we expect this ratio to be at the level of 
$10 - 20 \% $. In two cases, we observe a significant deviation from this expectation. 
This means that, in these cases, we need a large normalization constant for the cloud 
state responsible for direct production (which implies large $N_2$), in order to 
reproduce the observed asymmetry. This is a consequence of neglecting the asymmetry 
generated by indirect production, i.e., in these cases, the approximation 
$\Phi_I^{D^-}=\Phi_I^{D^+}$ is not a good one. A more complete discussion, taking 
this fact into account is presented in \cite{nosnafita}. Here, for the sake of the 
argument, we prefer to keep the calculation simple and keep our fits as they are, 
being implicit that the large $N_2$'s mimic the inclusion of indirect production.

In conclusion, we have shown that the MCM provides a good understanding of the charm 
production asymmetries in terms of a simple physical picture with few parameters.  
It connects the behavior 
of the asymmetries at large $x_F$ with the charm meson momentum 
distribution within the cloud state. We can explain why we 
observe asymmetries, why they are different for different beams and  we 
are led to the conclusion that they may vanish at very large $x_F$. This 
approach can easily be extendend to charm baryon production and also to 
$B$ production \cite{nosnafita}.

\acknowledgments
This work has been supported by CNPq and  FAPESP, under contract number 
2000/04422-7.

\end{document}